\documentclass[runningheads]{llncs}
\usepackage[T1]{fontenc}
\usepackage{booktabs}
\usepackage{multirow}
\usepackage{float}
\usepackage[table]{xcolor}
\usepackage{hyperref}

\newcommand\modelname{\texttt{LesionGen}}
\newcommand\modelnamemain{\texttt{LesionGen-R$\&$B}}
\newcommand\modelnamerichonly{\texttt{LesionGen-R}}
\newcommand\modelnamestatic{\texttt{LesionGen-S}}

\newcommand\ttidpm{\texttt{T2I-DPM}}

\usepackage{graphicx}
\begin{document}
\title{\modelname{}: A Concept-Guided Diffusion Model for Dermatology Image Synthesis}
\titlerunning{\modelname{} for Skin Lesion Synthesis}
%
\author{
Jamil Fayyad\inst{1}\thanks{Corresponding author} \and
Nourhan Bayasi\inst{2} \and
Ziyang Yu\inst{3} \and
Homayoun Najjaran\inst{1}
}

\authorrunning{J. Fayyad et al.}

\institute{
University of Victoria, Canada\\
\and
University of British Columbia, Canada
\and
University of Toronto, Canada\\
\email{jfayyad@uvic.ca}
}
\maketitle              
\begin{abstract}
Deep learning models for skin disease classification require large, diverse, and well-annotated datasets. However, such resources are often limited due to privacy concerns, high annotation costs, and insufficient demographic representation. While text-to-image diffusion probabilistic models (\ttidpm{}s) offer promise for medical data synthesis, their use in dermatology remains underexplored, largely due to the scarcity of rich textual descriptions in existing skin image datasets. In this work, we introduce \modelname{}, a clinically informed \ttidpm{} framework for dermatology image synthesis. Unlike prior methods that rely on simplistic disease labels, \modelname{} is trained on structured, concept-rich dermatological captions derived from expert annotations and pseudo-generated, concept-guided reports. By fine-tuning a pretrained diffusion model on these high-quality image-caption pairs, we enable the generation of realistic and diverse skin lesion images conditioned on meaningful dermatological descriptions. Our results demonstrate that models trained solely on our synthetic dataset achieve classification accuracy comparable to those trained on real images, with notable gains in worst-case subgroup performance. Code and data are available \href{https://github.com/jfayyad/LesionGen}{here}.

\end{abstract}
\section{Introduction}
Deep learning (DL) has revolutionized medical image analysis, offering unprecedented accuracy in disease detection ~\cite{Xie_Multidisease_MICCAI2024} and classification~\cite{bayasi2023continual,bayasi2021culprit,bayasi2024biaspruner,bayasi2024continual,fayyad2024empirical,elkhayat2025foundation}. However, these models are inherently data-hungry, requiring vast amounts of high-quality, diverse training images to generalize well~\cite{fayyad2023out,fayyad2024exploiting}. In dermatology, this need is particularly acute because skin disease classification relies on nuanced visual cues that can vary across demographics, lighting conditions, and disease progression. Yet, real-world medical datasets are often limited due to privacy concerns, ethical restrictions, and the high cost of expert annotations. The scarcity of diverse, labeled skin disease images presents a significant barrier to developing robust DL models.

To address this issue, synthetic data generation has emerged as a promising solution, with generative models offering a powerful tool for augmenting medical datasets. Among these, diffusion probabilistic models (\texttt{DPM}s)~\cite{ho2020denoising} have demonstrated outstanding capabilities in generating high-fidelity images across diverse domains, including image synthesis~\cite{jiang2023cola,yellapragada2024pathldm}, translation~\cite{kui2025med}, classification~\cite{prusty2024enhancing} and segmentation~\cite{wu2024medsegdiff}. Particularly noteworthy are text-to-image diffusion models (\ttidpm{}s), which leverage natural language descriptions to generate realistic images aligned with specific conditions, achieving state-of-the-art performance. For instance, de Wilde et al.~\cite{de2023medical} adapted pre-trained \ttidpm{}s to medical imaging through textual inversion, successfully generating diagnostically accurate images across a range of modalities. Similarly, Chen et al.~\cite{chen2024eyediff} introduced EyeDiff, a \ttidpm{} trained on a variety of ophthalmic datasets, demonstrating its effectiveness in generating images of rare eye diseases to address the critical issue of data imbalance in diagnostic models.

However, despite the promising potential of \ttidpm{}s in medical imaging, their application to dermatology faces a significant limitation: unlike other medical modalities, where images are typically accompanied by detailed textual reports ~\cite{wang2017chestx}, skin disease datasets often lack such structured descriptions. This absence of rich textual metadata makes it challenging to fully leverage \ttidpm{}s for data synthesis in dermatology. To date, only two studies have applied \ttidpm{}s to skin lesion synthesis~\cite{sagers2023augmenting,akrout2023diffusion}, but both relied on overly simplistic textual descriptions limited to disease labels; e.g., `\texttt{an image of $<$label-only$>$}', where \texttt{$<$label-only$>$} is replaced by the name of the corresponding skin disease. This approach is insufficient, as it fails to capture the rich, nuanced visual features that define each condition, such as variations in texture, color, shape, and progression. These features are critical for accurate lesion classification and diagnosis, and without them, the generated images lack the diversity and specificity needed to enhance model performance effectively.

In this work, we propose \modelname{}, a novel \ttidpm{} framework for concept-driven skin image synthesis, primarily aimed at improving worst-case performance across subgroups. Our method generates rich image–caption pairs by producing structured, clinically meaningful descriptions of dermatological images. Specifically, we employ two concept-based text generation strategies to create these captions: (1) a strategy based on expert dermatological descriptions, where each image is annotated with seven clinically relevant diagnostic attributes; and (2) a strategy using pseudo-generated dermatological descriptions, where a vision-language model is guided to produce detailed medical text conditioned on these dermatological concepts. These concept-grounded captions are paired with their corresponding images to form high-quality training data for \modelname{}, enabling conditional image generation. Our results demonstrate that models trained solely on this synthetic data achieve competitive classification accuracy, with notable improvements in worst-case subgroup performance.

\section{Methodology} \label{sec:methodology}
\begin{figure}[t]
    \centering
    \includegraphics[width=0.8\linewidth]{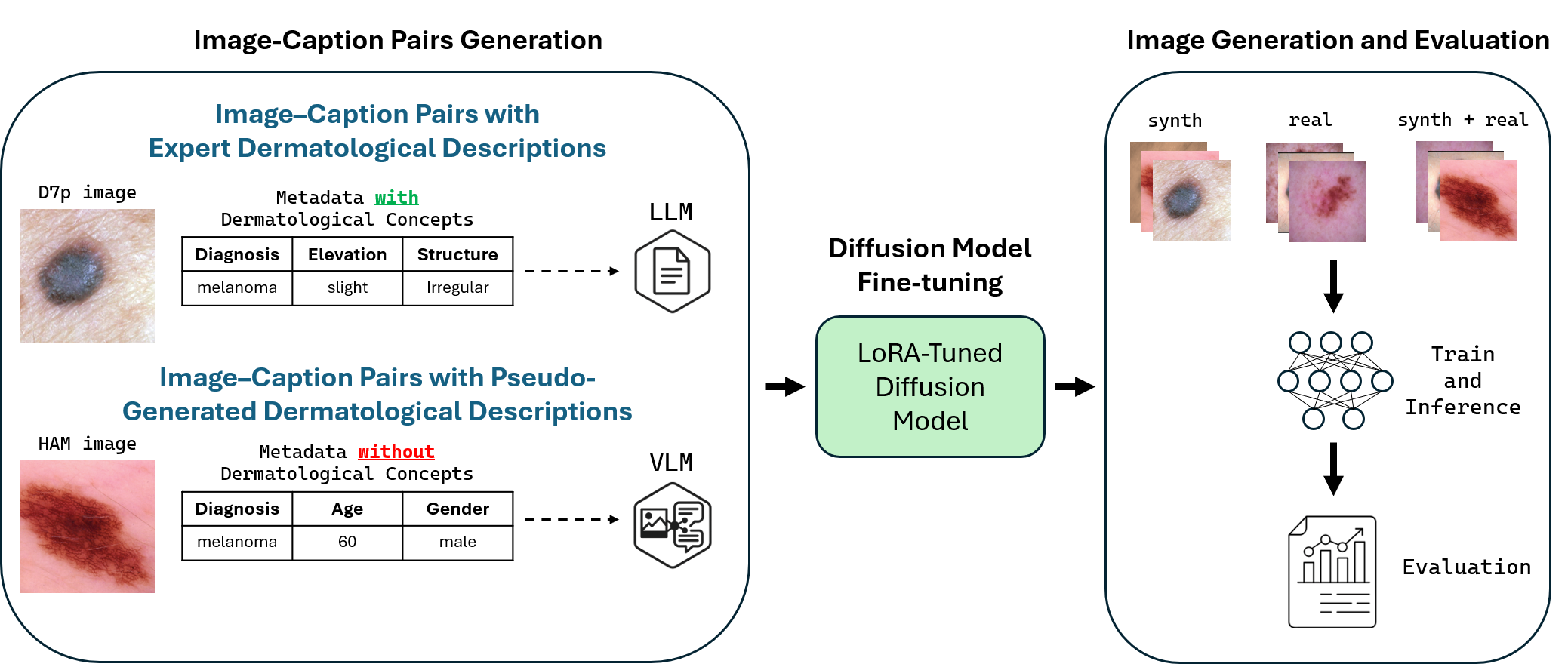}
        \caption{Overview of \modelname{}. Lesion images are paired with either expert dermatological descriptions transformed into text by an LLM, or pseudo-captions generated from image features using a VLM, forming a unified multimodal dataset. A LoRA-tuned diffusion model is trained on the multimodal dataset, and the trained model generates diverse, class-consistent images, which are then evaluated in the downstream classification task.}
    \label{fig:main_figure}
\end{figure}

We propose \modelname{}, a  dermatology image synthesis using a fine-tuned \ttidpm{}. As shown in Fig.~\ref{fig:main_figure}, the framework constructs image–caption pairs from expert annotations and pseudo-generated descriptions, then fine-tunes the diffusion model on these pairs for image generation and downstream evaluation.

\subsection{Training Data Construction} 
\begin{figure}[t]
    \centering
    \includegraphics[width=0.8\linewidth]{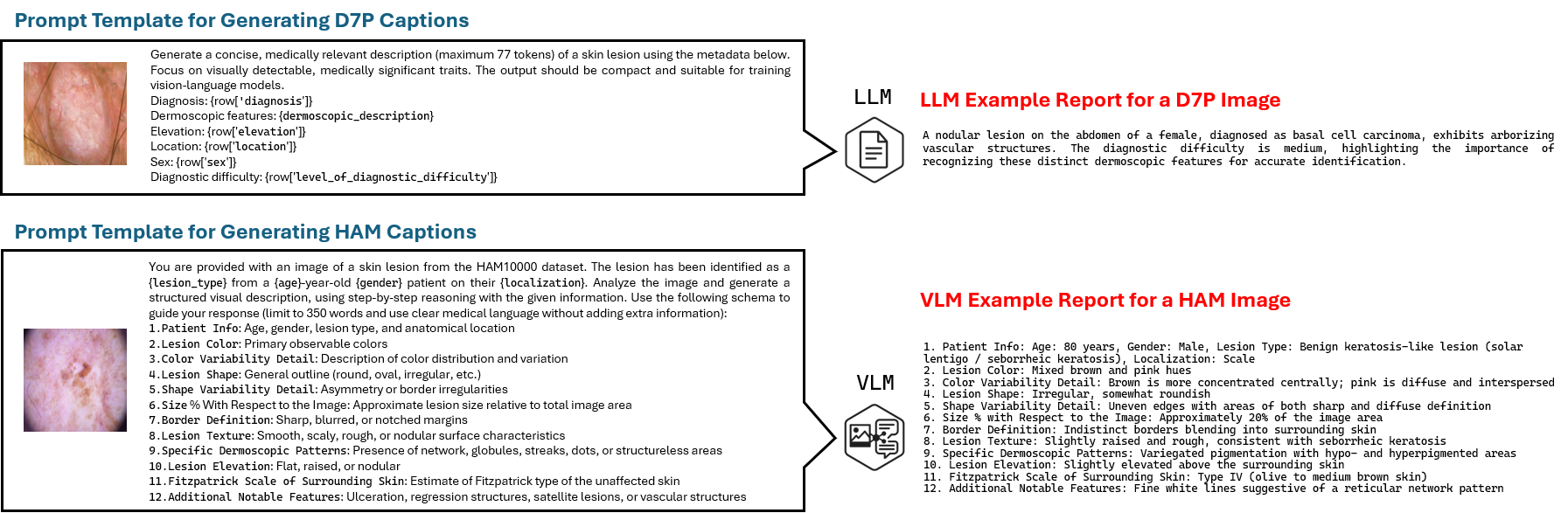}
        \caption{Prompt templates and example outputs for dermatological caption generation from (top) expert-annotated D7P data and (bottom) pseudo-generated HAM data.}
    \label{fig:example_reports}
\end{figure}

\noindent \textbf{Image–Caption Pairs with Expert Dermatological Descriptions.}  To construct clinically grounded  pairs, we use the D7P dataset~\cite{kawahara2018seven}, which contains 1,926 dermoscopic images spanning six diagnostic classes: nv, mel, bcc, df, bkl, and vasc. Each image is paired with structured metadata describing seven clinically meaningful attributes (referred to as \textit{concepts}), including pigmentation, lesion elevation, and structure. To transform this metadata into natural language, we prompt an LLM model with a structured template (Fig.~\ref{fig:example_reports}--top), generating dermatologist-style captions that serve as conditioning text for diffusion training.

\noindent \textbf{Image–Caption Pairs with Pseudo-Generated Dermatological  Descriptions.} The HAM10000 (HAM) dataset~\cite{ham} is a widely used benchmark in dermatology, comprising over 10,000 dermoscopic images across seven diagnostic classes (the six from D7P, plus akiec). Although HAM includes basic metadata, such as patient age, gender, lesion type, and anatomical location,  it lacks corresponding textual descriptions and, critically, does not provide annotations for key dermatological concepts, as in D7P, that are important to clinical diagnosis. To address this, we leverage a VLM model to generate structured dermatological descriptions that draw on both the available metadata (e.g., lesion label, patient age, gender) and the visual content of each image. Rather than prompting the VLM to produce free-form clinical descriptions, we craft targeted prompts that explicitly instruct the model to describe each image using the seven diagnostic concepts defined in the D7P dataset (e.g., pigmentation, elevation, structure). Although these attributes are not explicitly annotated in HAM, our prompting strategy enables the VLM to generate clinically meaningful, concept-based descriptions that  enhance the dataset’s utility for diffusion model fine-tuning. Fig.~\ref{fig:example_reports}--bottom illustrates the prompting strategy and an example output.


\subsection{Diffusion Model Fine-Tuning} 
We fine-tune a pretrained \texttt{DPM} using a combined set of expert and pseudo-generated image–caption pairs from D7P and HAM. The resulting model, \modelname{}, is trained to generate realistic and diverse skin lesion images conditioned on dermatological descriptions. To improve alignment between medical language and visual features, we leverage CLIP for text conditioning and apply Low-Rank Adaptation (LoRA) for efficient fine-tuning. Formally, \modelname{} samples an image $x_0$ given a text embedding $c$ as $x_0 \sim p_{\theta}(x_0 \mid c)$, where $c$ encodes the structured dermatological description. 


\section{Experiments and Results} 
We evaluate how synthetic skin lesion images generated by \modelname{} affect downstream classification performance. Specifically, we ask: (1) How effective are these synthetic images for training a CNN classifier (e.g., ResNet18~\cite{he2016deep})? and (2) How does prompt design influence the utility of generated images?

\subsection{Experimental Setup}
We fine-tune a pretrained Stable Diffusion v1.4 model~\cite{rombach2022high} on our multimodal dataset of image–caption pairs (see Section~\ref{sec:methodology}), which makes our \modelname{} model. In our main experimental setting, we use \textbf{rich and balanced prompts} as input to \modelname{} during image generation. We refer to this configuration as \modelnamemain{}. Rich refers to the prompts being identical to those used during diffusion model fine-tuning, whereas balanced refers to addressing class imbalance during image generation by ensuring each lesion class is associated with an equal number of prompts. Since some classes are underrepresented in the original dataset, we use GPT-4o to generate paraphrased versions of existing prompts for those classes. These paraphrases retain the original clinical meaning while introducing linguistic variation; e.g.,  turning `\texttt{A nodular melanoma featuring diffuse irregular pigmentation, irregular dots}' into `\texttt{A nodular melanoma showing uneven pigmentation, scattered dots}'. Using \modelnamemain{}, we generate a class-balanced synthetic dataset with 500 samples per class. A ResNet18  classifier is then trained on different combinations of the synthetic and real data, and evaluated on held-out real test sets from D7P and HAM.

\subsection{Baseline and Competing Method}
We compare \modelnamemain{} against two baselines. The first is a \textbf{real-only (upper bound)} configuration, where the ResNet18 classifier is trained solely on real images. The second is a prior SOTA method~\cite{akrout2023diffusion}, referred to as \textbf{p-SOTA}, in which the diffusion model is both \textbf{fine-tuned and sampled} using only static prompts that only include the label (e.g., \texttt{melanoma}) without any descriptive captions.

\subsection{Implementation Details}
\noindent \textbf{Prompt Generation.} We use GPT-4o in two modes: text-only for D7P (from structured metadata) and vision–language for HAM (using base64-encoded images with structured instructions). Prompts are generated with temperature 0.3 and a 77-token limit to comply with CLIP’s tokenization
limits.

\noindent \textbf{Diffusion Model Fine-Tuning.} 
We use Stable Diffusion v1.4, pretrained on the LAION-2B dataset \cite{schuhmann2022laion}, and fine-tune it using LoRA (Low-Rank Adaptation) on our image-caption dataset. The LoRA training uses a rank of 64, learning rate of 1e-5, and a constant scheduler. Training is performed for 15,000 steps using mixed precision (fp16), gradient checkpointing, and image augmentations like cropping and horizontal flipping. The output resolution is 256$\times$256. We qualitatively monitor progress by evaluating validation prompts every five epochs.

\noindent \textbf{Downstream Classification.} 
The downstream classifier is a ResNet18 trained from scratch. Input images are resized to 224$\times$224, normalized with a mean and standard deviation of 0.5, and randomly flipped during training. We use stochastic gradient descent with momentum 0.9, initial learning rate 0.01, and step decay (factor 0.1 every 10 epochs). Training uses a batch size of 32 and early stopping with patience of 5 epochs. Our code is implemented in PyTorch using the HuggingFace Diffusers library and OpenAI APIs.

\subsection{Main Results}
We train the ResNet18 model on a combined dataset consisting of up to 250 real images per class (when available), supplemented with synthetic images generated by \modelnamemain{} to reach a total of 500 training samples per class. This setup is referred to as (\textbf{synth+real}). The classification performance measured by overall accuracy and per-class precision on the D7P and HAM test sets is reported in Tables~\ref{tab:d7p_main} and~\ref{tab:ham_main}, respectively. 

On the \textbf{D7P test set} (Table~\ref{tab:d7p_main}), the classifier trained on \modelnamemain{}-generated data achieves the highest overall accuracy ({65.3\%}), slightly outperforming the real-only model ({65.0\%}) and significantly surpassing the p-SOTA baseline ({45.0\%}). Beyond overall accuracy, our method yields major improvements in worst-class performance. Notably, the precision for the {df} class increases from {0.000} (real-only) and {0.250} (p-SOTA) to {0.500} with \modelnamemain{}, demonstrating that our rich and balanced generation strategy enables the classifier to recover performance on previously underrepresented classes. Additionally, \modelnamemain{} provides the highest precision in 4 out of 6 classes, including {mel} and {nv}, which are critical in clinical diagnosis.

On the \textbf{HAM test set} (Table~\ref{tab:ham_main}), the classifier achieves the best overall accuracy ({75.6\%}) with our approach, outperforming the real-only  ({73.7\%}) and p-SOTA ({58.7\%}) scenarios. The improvements in rare or challenging classes are particularly notable: {df} improves from {0.000} (real-only) and {0.077} (p-SOTA) to {0.418} with our method, and {akiec} rises from {0.273} to {0.375}. Our method also leads to the best precision in 5 out of 7 classes, showing that \modelnamemain{}-generated data improves class-wise consistency without overfitting to majority classes. While {vasc} and {nv} show a slight drop compared to real-only, they remain strong overall, and this trade-off results in a more balanced and robust classifier.

\begin{table*}[t]
\centering
\caption{Classification performance on the D7P test set. We report overall accuracy and per-class precision. The best results are \textbf{bolded} in green-shaded cells.}
\label{tab:d7p_main}
\resizebox{\textwidth}{!}{%
\begin{tabular}{lccccccc}
\toprule
\textbf{Experiment} & \textbf{Overall Accuracy} & \textbf{bcc} & \textbf{bkl} & \textbf{df} & \textbf{mel} & \textbf{nv} & \textbf{vasc} \\
\midrule\midrule
p-SOTA  (synth+real)  & 0.450 & \cellcolor{green!20}\textbf{0.333} & 0.125 & 0.250 & 0.627 & 0.712 & 0.250 \\
{real-only}  & 0.650 & 0.214 & \cellcolor{green!20}\textbf{0.333} & 0.000 & 0.636 & 0.678 & 0.000 \\
\hline
\multicolumn{8}{c}{\textbf{Ours}} \\
\hline
\modelnamemain{} (synth+real) & \cellcolor{green!20}\textbf{0.653} & 0.188 & 0.316 &\cellcolor{green!20}\textbf{ 0.500 }& \cellcolor{green!20}\textbf{0.658} & \cellcolor{green!20}\textbf{0.721} & \cellcolor{green!20}\textbf{0.251} \\
\hline
\end{tabular}%
}
\end{table*}

\begin{table*}[t]
\centering
\caption{Classification performance on the HAM test set. We report overall accuracy and per-class precision. The best results are \textbf{bolded} in green-shaded cells.}
\label{tab:ham_main}
\resizebox{\textwidth}{!}{%
\begin{tabular}{lcccccccc}
\toprule
\textbf{Experiment} & \textbf{Overall Accuracy} & \textbf{akiec} & \textbf{bcc} & \textbf{bkl} & \textbf{df} & \textbf{mel} & \textbf{nv} & \textbf{vasc} \\
\midrule\midrule
{p-SOTA  (synth+real) } & 0.587 & 0.231 & 0.364 & 0.343 & 0.077 & 0.313 & \cellcolor{green!20} \textbf{0.954} & 0.434 \\
real-only                              & 0.737 & 0.273 & \cellcolor{green!20} \textbf{0.488} & 0.527 & 0.000 & 0.594 & 0.812 & \cellcolor{green!20} \textbf{1.000} \\
\hline
\multicolumn{9}{c}{\textbf{Ours}} \\
\hline
\modelnamemain{} (synth+real)     & \cellcolor{green!20} \textbf{0.756} & \cellcolor{green!20} \textbf{0.375 }& 0.421 & \cellcolor{green!20} \textbf{0.538} & \cellcolor{green!20} \textbf{0.418} & \cellcolor{green!20} \textbf{0.622} & 0.807 & 0.315 \\ \hline
\end{tabular}%
}
\end{table*}

\begin{table*}[t]
\centering
\caption{Ablation study results on the D7P test set, reporting overall accuracy and per-class precision.  The best results in each ablation are \textbf{bolded} in green-shaded cells. }
\label{tab:d7p_ablation}
\resizebox{\textwidth}{!}{%
\begin{tabular}{lccccccc}
\toprule
\textbf{Experiment} & \textbf{Overall Accuracy} & \textbf{bcc} & \textbf{bkl} & \textbf{df} & \textbf{mel} & \textbf{nv} & \textbf{vasc} \\
\hline
\multicolumn{8}{c}{\textbf{Ablation A: Impact of Using Synthetic Data Alone}} \\
\hline
p-SOTA (synth only)      & 0.446 & 0.056 & 0.113 & 0.000 & 0.636 & \cellcolor{green!20}\textbf{0.715} & \cellcolor{green!20}\textbf{0.111 }\\ 
\modelnamemain{} (synth only)      & \cellcolor{green!20}\textbf{0.551} & \cellcolor{green!20}\textbf{0.100} & \cellcolor{green!20}\textbf{0.143} & 0.000 & \cellcolor{green!20}\textbf{0.652} & 0.690 & 0.056 \\ 
\hline
\multicolumn{8}{c}{\textbf{Ablation B: Impact of Removing Prompt Balancing
}} \\
\hline
\modelnamerichonly{} (synth only) & 0.577 & 0.000 & \cellcolor{green!20}\textbf{0.250} & \cellcolor{green!20}\textbf{0.059 }& \cellcolor{green!20}\textbf{0.623} & 0.651 & 0.050 \\
\modelnamerichonly{} (synth+real) & \cellcolor{green!20}\textbf{0.611} & \cellcolor{green!20}\textbf{0.250} & 0.067 & 0.000 & 0.596 & \cellcolor{green!20}\textbf{0.670} & \cellcolor{green!20}\textbf{0.125} \\ 
\hline 
\multicolumn{8}{c}{\textbf{Ablation C: Impact of Using Static, Label-Only Prompts}} \\
\hline
\modelnamestatic{} (synth only) & 0.324 & 0.000 & 0.047 & 0.028 & 0.280 & 0.692 & 0.067 \\
\modelnamestatic{} (synth+real) &\cellcolor{green!20}\textbf{ 0.637} & \cellcolor{green!20}\textbf{0.077} & \cellcolor{green!20}\textbf{0.222} & \cellcolor{green!20}\textbf{0.300} & \cellcolor{green!20}\textbf{0.597} & \cellcolor{green!20}\textbf{0.733} & \cellcolor{green!20}\textbf{0.300 }\\ \hline 
\end{tabular}%
}
\end{table*}

\begin{table*}[t]
\centering
\caption{Ablation study results on the HAM test set, reporting overall accuracy and per-class precision. The best results in each ablation are \textbf{bolded} in green-shaded cells.}
\label{tab:ham_ablation}
\resizebox{\textwidth}{!}{%
\begin{tabular}{lcccccccc}
\toprule
\textbf{Experiment} & \textbf{Overall Accuracy} & \textbf{akiec} & \textbf{bcc} & \textbf{bkl} & \textbf{df} & \textbf{mel} & \textbf{nv} & \textbf{vasc} \\
\hline 
\multicolumn{9}{c}{\textbf{Ablation A: Impact of Using Synthetic Data Alone}} \\
\hline
p-SOTA (synth only)      & 0.315 & 0.009 & 0.126 & 0.130 & 0.000 & 0.236 & \cellcolor{green!20}\textbf{0.978} & \cellcolor{green!20}\textbf{0.091}  \\ 
\modelnamemain{} (synth only) & \cellcolor{green!20}\textbf{0.524 }& \cellcolor{green!20}\textbf{0.145} & \cellcolor{green!20}\textbf{0.188} & \cellcolor{green!20}\textbf{0.210} & \cellcolor{green!20}\textbf{0.045} & \cellcolor{green!20}\textbf{0.489} & 0.681 & 0.080 \\ 
\hline
\multicolumn{9}{c}{\textbf{Ablation B: Impact of Removing Prompt Balancing
}} \\
\hline
\modelnamerichonly{} (synth only) & 0.428 & 0.000 & 0.078 & 0.169 & 0.000 & 0.164 & 0.927 & \cellcolor{green!20}\textbf{1.000} \\
\modelnamerichonly{} (synth+real) &\cellcolor{green!20} \textbf{0.585} & \cellcolor{green!20}\textbf{0.230} & \cellcolor{green!20}\textbf{0.389 }& \cellcolor{green!20}\textbf{0.312} & 0.000 & \cellcolor{green!20}\textbf{0.298} & \cellcolor{green!20}\textbf{0.938} & 0.475 \\ 
\hline 
\multicolumn{9}{c}{\textbf{Ablation C: Impact of Using Static, Label-Only Prompts
}} \\
\hline
\modelnamestatic{} (synth only) &  0.176 & 0.011 & 0.211 & 0.068 & 0.000 & 0.214 & \cellcolor{green!20}\textbf{1.000} & 0.000 \\
\modelnamestatic{} (synth+real) &  \cellcolor{green!20}\textbf{0.608} & \cellcolor{green!20}\textbf{0.239} & \cellcolor{green!20}\textbf{0.370 }& \cellcolor{green!20}\textbf{0.405} & \cellcolor{green!20}\textbf{0.065} & \cellcolor{green!20}\textbf{0.326} & 0.942 & \cellcolor{green!20}\textbf{0.512} \\ \hline
\end{tabular}%
}
\end{table*}
\subsection{Ablation Studies}
We conduct ablation studies to answer three questions: (A) Can synthetic data alone yield strong performance? (B) What is the impact of prompt balancing on class-wise and overall accuracy? and (C) How much does prompt enrichment improve results over static, label-only prompts?

\noindent \textbf{Ablation A: Impact of Using Synthetic Data Alone.}  
In ablation study A, we remove all real training images and train the ResNet18 model solely on synthetic data generated by either p-SOTA or \modelnamemain{}. As shown in Tables~\ref{tab:d7p_ablation} and~\ref{tab:ham_ablation}, this leads to a clear drop in both overall accuracy and worst-class precision compared to the synth+real setting (Tables~\ref{tab:d7p_main} and~\ref{tab:ham_main}). For instance, on the HAM dataset, training on p-SOTA synthetic images results in just {31.5\%} accuracy, whereas synthetic images from \modelnamemain{} yield a significantly higher {52.4\%}. While performance still falls short of configurations that include real data, these results indicate that our concept-driven generation approach produces higher-quality and more informative samples than prior methods. Overall, this ablation confirms that synthetic data alone is not sufficient, but can be highly effective when used in combination with real samples. 

\noindent \textbf{Ablation B: Impact of Removing Prompt Balancing.}  
In this ablation, we generate synthetic data using rich, concept-guided prompts but without applying prompt balancing for underrepresented classes. As shown in Tables~\ref{tab:d7p_ablation} and~\ref{tab:ham_ablation}, removing balancing results in a decline in overall accuracy and highly inconsistent class-wise performance, particularly when compared to the full \modelnamemain{} setting (Tables~\ref{tab:d7p_main} and~\ref{tab:ham_main}) where balancing was applied. Furthermore, the classifier performs well on majority classes; e.g., on HAM, precision for {nv} reaches {93.8\%} in the synth+real setting, and {vasc} reaches {100\%} even with synthetic data alone. However, performance on minority classes could collapse entirely; e.g., df. 

\noindent \textbf{Ablation C: Impact of Using Static, Label-Only Prompts.} 
Finally, we test the necessity of the rich prompts by replacing them with static, label-only prompts. In the synth-only setting, performance deteriorates markedly,
with accuracy dropping to {32.4\%} on D7P (Table~\ref{tab:d7p_ablation}) and {17.6\%} on HAM (Table~\ref{tab:ham_ablation}), with multiple classes exhibiting zero or near-zero precision. Even in the synth+real setting, performance lags behind the rich and balanced prompts configuration (Tables~\ref{tab:d7p_main} and~\ref{tab:ham_main}), demonstrating that label-only prompts lack the semantic richness needed to guide the diffusion model toward clinically meaningful generation.

\subsection{Visualization Results}
\begin{figure}[t]
    \centering
    \includegraphics[width=0.8\linewidth]{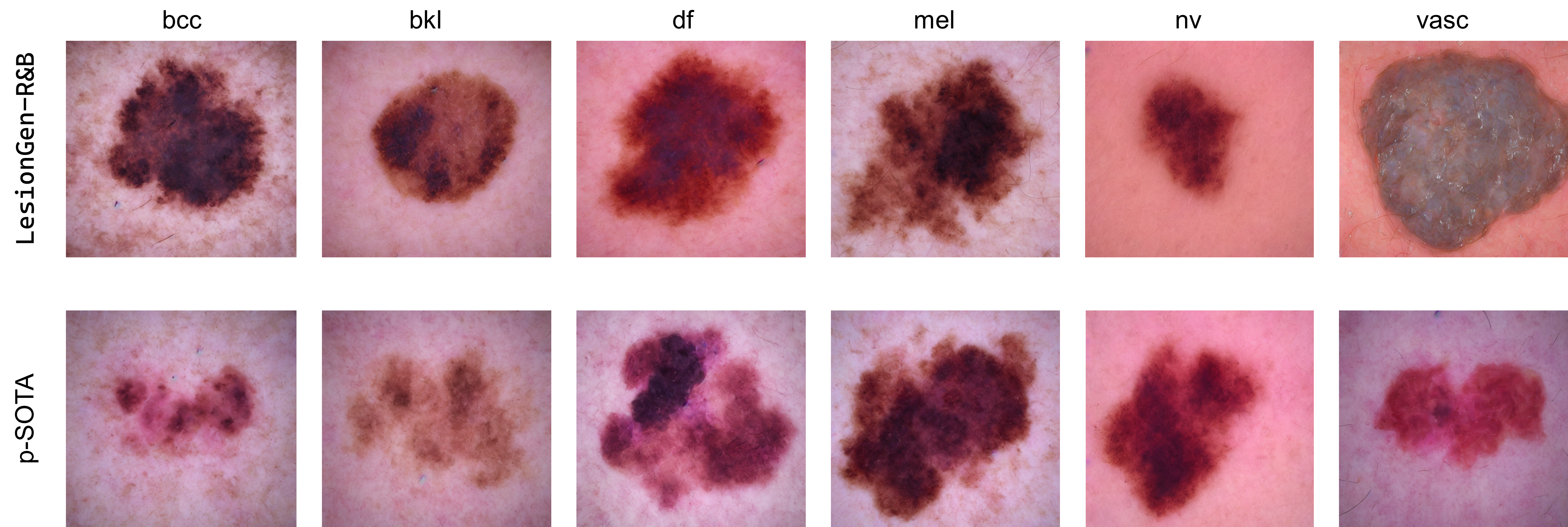}
\caption{Examples of synthetic skin lesion images generated from \modelnamemain{} (top) and p-SOTA (bottom) across six classes.}
    \label{fig:example_synth}
\end{figure}

Fig.~\ref{fig:example_synth} shows samples from \modelnamemain{} (top) and p-SOTA (bottom). Despite visual similarity, our method’s superior classification performance suggests it captures subtle, clinically relevant features beyond human perception.

\section{Conclusion}
In this work, we demonstrate the effectiveness of combining text-to-image diffusion models with concept-guided dermatological prompts for generating high-quality synthetic skin lesion images. Unlike prior approaches that rely on simple label-based conditioning, our method leverages rich, structured descriptions aligned with clinical concepts, and balances class representation through prompt paraphrasing. This design enables the generation of semantically diverse and class-balanced datasets that complement real-world dermatology benchmarks. Training a ResNet18 on \modelname{}’s outputs significantly boosts classification performance, especially for underrepresented groups, outperforming the baselines. Future work includes expanding to skin tone diversity, interactive refinement, and multi-modal conditioning.

 \bibliographystyle{splncs04}
 \bibliography{mybibliography}
 
\end{document}